\documentclass[twocolumn,showpacs,amssymb,prb,aps,superscriptaddress]{revtex4}
\usepackage[latin1]{inputenc}
\usepackage{graphicx}
\usepackage{color}
\usepackage{soul}

\def\be{\begin{equation}}
\def\ee{\end{equation}}
\begin{document}

\title{\bf Spin density wave instabilities in the NbS$_2$ monolayer}
\author{F. G\"uller}
\affiliation{Centro At\'{o}mico Constituyentes, GIyANN, CNEA, Av. Gral. Paz 1499, San Mart\'{i}n, Buenos Aires, Argentina}
\affiliation{Consejo Nacional de Investigaciones Cient\'{i}ficas y T\'{e}cnicas, Av. Rivadavia 1917 (C1033AAJ), Buenos Aires, Argentina}
\author{V. L. Vildosola}
\affiliation{Centro At\'{o}mico Constituyentes, GIyANN, CNEA, Av. Gral. Paz 1499, San Mart\'{i}n, Buenos Aires, Argentina}
\affiliation{Consejo Nacional de Investigaciones Cient\'{i}ficas y T\'{e}cnicas, Av. Rivadavia 1917 (C1033AAJ), Buenos Aires, Argentina}
\author{A. M. Llois}
\affiliation{Centro At\'{o}mico Constituyentes, GIyANN, CNEA, Av. Gral. Paz 1499, San Mart\'{i}n, Buenos Aires, Argentina}
\affiliation{Consejo Nacional de Investigaciones Cient\'{i}ficas y T\'{e}cnicas, Av. Rivadavia 1917 (C1033AAJ), Buenos Aires, Argentina}
\affiliation{Departamento de F\'{i}sica Juan Jos\'{e} Giambiagi, FCEyN-UBA, Intendente Güiraldes 2160 (C1428EGA), Buenos Aires, Argentina}

\begin{abstract}
In the present work, we study the magnetic properties of the NbS$_2$ monolayer by first-principles calculations.
The transition metal dichalcogenides (TMDC) are a family of laminar materials presenting exciting properties such as
charge density waves (CDW), superconductivity and metal-insulating transitions among others. 2H-NbS$_2$ is a particular case
within the family, because it is the only one that is superconductor without exhibiting a CDW order.
Although no long range magnetic order was experimentally observed in the TMDC, we show here that 
the single monolayer of NbS$_2$ is on the verge of a spin density wave (SDW) phase.
Our calculations indicate that a wave-like magnetic order is stabilized in the NbS$_2$ monolayer in the
presence of magnetic defects or within zig-zag nanoribbons, due to the presence of unpaired electrons. We calculate the real
part of the bare electronic susceptibilty and the corresponding nesting function of the clean NbS$_2$ monolayer,
showing that there are strong electronic instabilities at the same wavevector asociated with the calculated SDWs, also
corresponding with one of the main nesting vectors of the Fermi surface. We conclude that the physical mechanism
behind the spin-wave instabilities are the nesting properties, accentuated by the quasi 2D character of this system, and the rather strong Coulomb interactions of the 4d band of the Nb atom. We also estimate the amplitude of the spin-fluctuations and find that they are rather large, as expected for a system on the verge of a quantum critical transition.
\end{abstract}

\date{\today}

%75.30.Fv Spin-density waves
%75.75.-c Magnetic properties of nanostructures
%68.65.-k Low-dimensional, mesoscopic, nanoscale and other related systems: structure and nonelectronic properties
%73.22.-f Electronic structure of nanoscale materials and related systems

\maketitle

\section{Introduction \label{sec:Intro}}
Transition metal dichalcogenides (TMDC), MX$_2$ (X = S,Se or Te), constitute a family of compounds with fascinanting physical properties such as charge ordered phases and superconductivity. They are characterized by a laminar structure similar to that of graphite. Each monolayer is actually a trilayer, composed by a plane of metal atoms sandwiched by two planes of S, Se or Te. Within one trilayer, the atoms are covalently bound while separate trilayers are held together mainly through weak van der Waals interactions.
In pristine TMDC there are two different bulk structures (polytypes), abbreviated by 1T and 2H, corresponding to the either octahedral or trigonal prismatic coordination of the M atom. The primitive cell of the 1T and 2H polytypes contains one and two monolayers, respectively.
%A single, isolated monolayer of the 2H polytype is abbreviated by 1H.

Bulk TMDC can present strikingly different behaviours dependending on the polytype. For instance, charge density wave (CDW) transitions occur in both 1T and 2H  structures but with very different structural and electronic properties. Some 1T systems are Mott insulators while in the 2H family there is no trace of Mott physics. 
In particular, the 2H-MX$_2$ can be either metallic (such as Nb(S,Se)$_2$ and Ta(S,Se)$_2$) or band insulators (as MoS$_2$ and  WS$_2$) depending on the filling of 
the $d$ band. Notably, also both polytypes can present a superconducting state. Many works have been devoted to studying the driving force of the CDW and its relation with superconductivity (SC). Naturally, the quasi two-dimensional structure of these laminar materials is prone to a nesting scenario characterized by paralell pieces of the Fermi surface at a given "nesting" vector, $\mathbf{q_n}$. The traditional understanding would be that under this nesting condition, the electron system might become unstable and induce a  CDW  transition in the Peierls- like manner and/or a SDW one.
However, it has been shown for 2H-NbSe$_2$ and 2H-TaSe$_2$, two prototypes of CDW systems, both by angle resolved photemission experiments (ARPES) 
and ab-initio calculations, that the charge ordering wavevector, q$_{CDW}$ is different from q$_n$,  indicating that a simple nesting 
model\cite{Johannes2006,Johannes2008} cannot account for the CDW in these materials. 
Instead, an enhancement of the electron-phonon coupling at q$_{CDW}$ has been proposed to be the driving force.
More recently, ARPES data supported by theoretical calculations of the k-resolved susceptibility, suggested that the CDW instability
is dominated by finite energy trasitions from states far away from the Fermi surface. In any case, the general consensus is that the CDW
in these systems is not of pure electronic origin \cite{Laverock2013}.

Among the TDMC, 2H-NbS$_2$ is a particular case, because it is a superconductor like 2H-NbSe$_2$, with a similar T$_c$, but no
CDW ordering has been observed experimentally. First-principles calculations have shown that the CDW is suppressed in this material 
due to anharmonic effects\cite{Leroux2012}. These results suggest that being the superconducting properties very similar to 
those of the isolectronic 2H-NbSe$_2$, either the nature of the ordered phase is not relevant for the superconducting state or 
there are other types of instabilities unexplored up to now. As was recently reviewed in Refs. \onlinecite{Klemm,Hirsch}, the TMDCs present 
many properties that are similar to the ones observed in the unconventional cuprates and Fe-based superconductors; a pseudogap behaviour is one of them.
However, no SDW nor any other kind of long range magnetic order has been observed in any of these TMDCs until now.

Experimentally, exfoliation techniques used to obtain graphene have been adjusted to TMDC to produce analogous samples of 
lower dimansionality, being it few monolayers, a single monolayer  and even nanoribbons or flakes. Another way to lower 
the dimensionality in TMDC is through the intercalation with organic molecules or transition metals.
Interestingly, it has been very recently observed, through optical and electrical transport measurements, 
that CDW effects in NbSe$_2$ are strongly enhanced in the monolayer\cite{Nature2015}.

The fact that Nb is a 4d transition metal with important electronic exchange interactions and the feasibility to lower the 
dimensionality of this type of laminar systems, motivated us to study the electronic properties of the NbS$_2$ monolayer. 
The absence of CDW order makes NbS$_2$ particularly appealing due to this relative simplicity.

In a previous work, we have shown by means of ab-initio calculations that NbS$_2$ zig-zag nanoribbons of different widths 
develop a wave-like pattern in their magnetic moments\cite{Guller2013}. In the present work, we generalize the study of 
the magnetic instabilities and gain insight into the physical mechanism that drives the observed magnetic solutions.
By  calculating  the real part of the bare electronic susceptibility in the constant matrix element approximation\cite{Chan1973}, 
and the corresponding nesting function we show that the nesting properties of the NbS$_2$ monolayer, together 
with the rather strong Coulomb exchange interaction of the 4d band, put this system to be on the verge of SDW transition. 
As expected for a system close to a quantum critical magnetic transition, there should exist large spin fluctuations. 
We have estimated the amplitude of these fluctuations through the fluctuation-dissipation theorem obtaining a value that is consistent with this scenario. We reinforce this picture by showing that doping, defects or ribbon edges can stabilize spin density waves in NbS$_2$ monolayers. 

The paper is organized as follows. In section \ref{sec:Comp} we describe the computational details. In section \ref{sec:Results} we study the 
electronic properties of the NbS$_2$ monolayer and the magnetic solutions that are obtained when placing defects or creating ribbon edges. We then analyze the static bare susceptibility of the monolayer and show the wave-like magnetic patterns to be SDW states. The estimation of the spin-fluctuations of this system are described at the end of this section. Finally, we present our conclusions in section \ref{sec:Conclusions}.

\section{Computational details\label{sec:Comp}}

Previous works have shown that \textit{ab initio} calculations based on density functional theory (DFT) correctly describe the CDW phase
in several 2H-MX$_2$ (see for example Refs. \onlinecite{Calandra2009}-\onlinecite{Sharma2002}). In this work, to study the magnetic properties of the NbS$_2$ monoloyer, the calculations were preformed using both the VASP \cite{Kresse1993,Kresse1996,VASPspecific} and WIEN2k \cite{WIEN2k,WIEN2kspecific} codes. The cross-check is particularly necessary  in the case of the pristine NbS$_2$ monoloyer since, as it will be shown below, there are several competing phases very close in energy. For the exchange correlation potential the generalized gradient approximation as parametrized by Perdew \textit{et al} \cite{Perdew1996} was employed. The atom positions were modified until all forces were smaller than 0.02 eV/\AA{}{}. To model the monolayers, a supercell with 13 \AA{}  vacuum in the perpendicular direction was used in order to avoid interactions between monolayers since, periodic boundary conditions were applied in all directions.
The k-point grid in the first Brillouin zone was 23x23x1. For calculations of the Fermi surface and susceptibility it was enlarged to 200x200x1, and a 8 meV Fermi temperature smearing was used. Only the VASP code was employed to analyze nanoribbons, which were simulated using a supercell with a 20 \AA{}  vacuum region in the two perpendicular directions. The k-point grid in these cases was 35x1x1. We perform spin polarized (SP) and non SP calculations, with and without considering the spin-orbit interaction (SO), and discuss its effects in the results obtained.

\section{Results\label{sec:Results}}

To our knowledge, the experimental structural parameters of the hexagonal NbS$_2$ monolayer have not been reported yet. 
The available data corresponds to bulk samples\cite{Raybaud1997}. The obtained structural parameters, \textit{a} and \textit{zc} (being \textit{zc} the height of the S atom relative to the Nb plane), of the relaxed NbS$_2$ monolayer are very close to the ones measured in the bulk. The relaxed monolayer has a slightly larger lattice constant $\textit{a}$ and a small increase of $\textit{zc}$. The obtained values are $a=3.34$ \AA{}  and $zc=1.56$ \AA{}, which are within 1\% and 5\% of the experimental ones for the bulk, respectively\cite{NotaZC}. %These results are in line with previous studies (see for example Ref. \onlinecite{Ataca2012}).

\begin{figure}[!tbh]
\centering
\includegraphics[width=4.5cm]{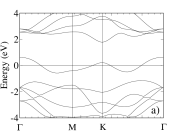} 
\includegraphics[width=3.5cm]{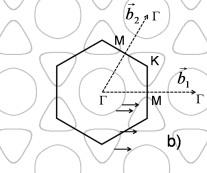} 
\caption{\label{fig:BandStructure} \textbf{a)} Non SP band structure of the NbS$_2$ monolayer, without SO. The single band crossing the Fermi level (zero energy) is primarily of Nb-$d_{z^2}$ character. \textbf{b)} Corresponding Fermi surface of the NbS$_2$ monolayer. The first Brillouin zone is drawn in the centre of the plot. $\mathbf{b_1}$ and $\mathbf{b_2}$ are the reciprocal lattice primitive vectors. One of the main nesting vectors is indicated by arrows.}
\end{figure}

The most stable solution for the NbS$_2$ monolayer, both with and without SO coupling, is the non SP one, if the calculations are carried out in the primitive cell (1x1). There is also a ferromagnetic phase, very close in energy, with 0.3 $\mu$B/Nb and 0.4 $\mu$B/Nb, with and without SO, respectively. 
The corresponding bandstructure (without SO) is shown in Fig. \ref{fig:BandStructure}a. There are six full S p-type bands, 
one band crossing the Fermi level of mainly Nb-$d_{z^2}$ character (A´$_1$ of the $D_{3h}$ group) and four empty 
Nb-d bands with E´ and E´´ symmetries. Due to the strong covalency of this material, there is a strong p-d hybridization 
effect between the full and empty bands. In Fig. \ref{fig:BandStructure}b we show the corresponding Fermi surface and the 
first Brillouin zone. One of the main nesting vectors is in the $\Gamma$M direction, indicated by arrows joining parts of the Fermi surface.

\begin{figure}[thb]
\centering
\includegraphics[width=8.3cm]{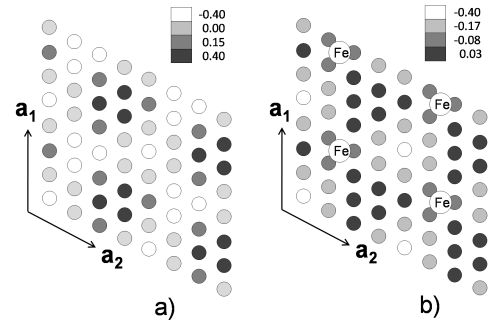} 
\caption{\label{fig:2DPattern} Schematic representation of the magnetic moments (without SO) of the Nb atoms in a 4x4 supercell. All values, in $\mu_B$, are approximate and intended as a guide for the eyes only. \textbf{a)} Wave-like configuration that competes in energy with the non SP solution in pristine NbS$_2$. \textbf{b)} Configuration obtained with a single Fe adsorption. The Fe atom has a magnetic moment of -2.6 $\mu_B$. $\mathbf{a_1}$ and $\mathbf{a_2}$ are the lattice prmitive vectors and the 4x4 supercell is repeated twice along each direction.}
\end{figure}

When a larger supercell is considered, non-trivial magnetic configurations appear. We study here a 4x4 supercell and observe
 that a wave-like pattern in the magnetic moments of the Nb atoms, shown schematically in Fig. \ref{fig:2DPattern}a, 
competes in energy with the non SP solution. When using the Wien2k code, without SO, the obtained wave-like configuration 
is energetically more favourable (by 2 meV/Nb) than the non SP case. The maximum value of the magnetic moments is 
0.4 $\mu_B$/Nb. However, when the SO coupling is included in the calculation, the non SP solution becomes more stable than 
the wave-like magnetic one with an energy difference of 4 meV/Nb. The energetics obtained with VASP is similar except for 
the fact that one can also find an intermediate, nearly degenerate, ferromagnetic state, less than 1 meV/Nb from the non SP one. The magnetic moments in the ferromagnetic case are 0.16 $\mu_B$/Nb. It is important to remark that all the solutions are very close in energy and the values of the magnetic moments are highly dependent on the atomic positions and lattice constants. Also, the shape of the magnetic patterns is constrained by the periodic boundary condition of the 4x4 supercell.

The existence of several magnetic configurations competing in energy and the high sensitivity to structural parameters, point towards a system with a very large magnetic susceptibility. These wave-like solutions can be fully stabilized if a chemical perturbation is introduced. As an example, we show in Fig. \ref{fig:2DPattern}b a schematic representation of the Nb magnetic pattern obtained for a 4x4 supercell of NbS$_2$ with a single Fe atom adsorbed on the monolayer. Similar solutions are obtained by substitutional doping with, for instance, Co or Fe, vacancies or through other adsorbed species. Small changes in the atomic positions or lattice constants, or the introduction of the spin-orbit interaction, do not destroy the magnetic patterns.
% provided the defect introduces unpaired bonds or a magnetic moment in the monolayer.
The configuration is similar to the one in Fig. \ref{fig:2DPattern}a, corresponding to the pristine monolayer. 
However, unlike the later case, the wave-like configurations are energetically far more favoured than the non SP solution, especially if the defect introduces unpaired bonds or a magnetic moment in the monolayer. For example, the energy diferences in the 4x4 supercell are of around 5 meV/Nb for a substitution
of Nb by Fe or Co, of 25 meV/Nb for an Nb vacancy and 38 meV/Nb for the Fe adsorbtion in Fig. \ref{fig:2DPattern}b.
% in all cases larger than the 2 meV/Nb of the pristine monolayer. 

%To estimate the energy difference, we first calculate the total energies with and without spin polarization. We then substract the contribution from the S atoms around the vacancy itself. This can be approximated as the difference in the total energies obtained from the non SP solution and a SP calculation with the magnetic moments of all Nb atoms in the supercell fixed to zero. In this way, the estimated energy of the wave-like solution with respect to the non SP case is approximately 25 meV/Nb, an order of magnitude larger than in the pristine monolayer.  

\begin{figure*}[!thb]
\centering
\includegraphics[width=14cm]{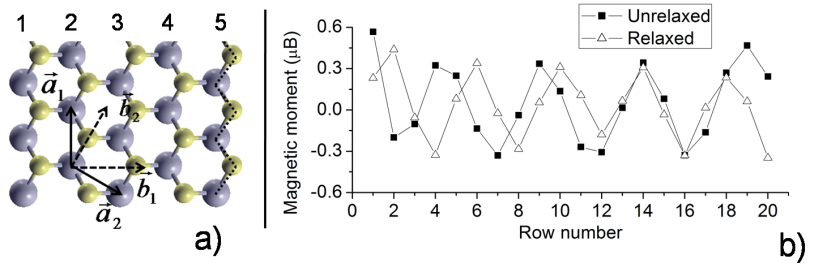} 
\caption{\label{fig:RigidRibbonMag} \textbf{a)} Top view of the NbS$_2$ zig-zag ribbon structure. Large spheres stand for transition metal atoms, small spheres for chalcogen atoms. The real space ($\mathbf{a_{1,2}}$) and reciprocal space ($\mathbf{b_{1,2}}$) lattice vectors of the NbS$_2$ monolayer are explicitly plotted. The ribbons are periodic in the $\mathbf{a_1}$ direction and the corresponding edge is marked by the dashed line. Rows of Nb atoms are numbered consecutively, starting on the Nb terminated edge. \textbf{b)} Magnetic moments of the individual Nb atoms across the unrelaxed and relaxed NbS$_2$ zig-zag ribbons of width N=20, without SO.}
\end{figure*}

If the hexagonal symmetry of the monolayer is broken by making an infinite line of defects or considering quasi-1D ribbons, the magnetic oscillations occur along a single direction. In Fig. \ref{fig:RigidRibbonMag}b, we show the obtained magnetic 
moments of the individual Nb atoms across NbS$_2$ zig-zag ribbons, in the relaxed and unrelaxed structures of equal 
width. The zig-zag cut of the ribbon is presented in Fig. \ref{fig:RigidRibbonMag}a; on top, the lattice primitive 
vectors of the monolayer in real and reciprocal space are shown. If the zig-zag edge is taken paralell to $\mathbf{a_1}$, 
then the width N is defined as the number of formula units in the ribbon primitive cell in the direction perpendicular 
to $\mathbf{a_1}$. Successive rows of Nb atoms are then numbered starting on the Nb terminated edge. The magnetic 
moments in Fig. \ref{fig:RigidRibbonMag}b correspond to ribbons with N=20. The wavy pattern is also present in ribbons of 
smaller width (shown in Ref. \cite{Guller2013}), however the periodic character is clearer for large N due to a less important finite size effect.  As in the case of the monolayers with point defects, these magnetic solutions are very robust and 
appear in both the relaxed and unrelaxed cases. Despite their low dimensionality, the NbS$_2$ ribbons are structurally very rigid. Relaxation affects mainly the positions of the edge atoms (rows 1 and 20), changing their magnetization. 
The difference in the wavelenght of the pattern is due to the slightly smaller lattice constant of the relaxed ribbon. 
When the SO coupling is considered, the oscillations have a slightly larger amplitude. We note that, as in the monolayer 
case, this pattern is stabilized by a magnetic perturbation, namely, the ribbon edges. If the edge magnetism is removed, 
the wave-like configuration disappears. In order to confirm this, we performed calculations on ribbons with edges 
passivated by hydrogen (not shown). Similarly to the better known case of MoS$_2$ zig-zag ribbons\cite{Botello2009}, 
H-passivation of all dangling bonds removes the edge magnetism and, in the NbS$_2$ ribbon, the wave-like pattern as well.

The fact that the wave-like magnetic patterns are present in the unrelaxed perturbated systems is indicating that 
the magnetic state is of purely electronic origin. We claim that these magnetic patterns are actually SDW instabilities 
of the NbS$_2$ monolayer, that are stabilized by magnetic point defects or ribbon edges. In order to get insight into 
the microspcopic origin of these magnetic instabilities, we calculate the real part of the bare electronic susceptibility 
in the static limit and constant matrix approximation \cite{Chan1973}, $\chi'_0(\mathbf{q})$, and the nesting function, 
N$_f(\mathbf{q})$, for the NbS$_2$ monolayer. The real part of the susceptibility reads 
$\chi'_0(\mathbf{q})=\sum_{\mathbf{k}}{\frac{f_{\mathbf{k}+\mathbf{q}}-f_{\mathbf{k}}}{\epsilon_{\mathbf{k}}-\epsilon_{\mathbf{k}+\mathbf{q}}}}$, where $f_{\mathbf{q}}$ is the Fermi function and $\epsilon_{\mathbf{k}}$ is the band energy. 
In the case of the monolayer, only one band crosses the Fermi level, so that the band index is omitted. The interband 
contributions in this case does not affect the shape of $\chi'_0(\mathbf{q})$. 

In the simplest case, where there is no periodic lattice distortion, the SDW becomes stable when the criterion $\chi'_0(\mathbf{q})\cdot V_{\mathbf{q}} > 1$ is met, being $V_{\mathbf{q}}$ the exchange matrix element interaction in the local 
approximation \cite{Chan1973}. The orientation of the SDW is given by the direction of $\mathbf{q}$ in real space, and its' period 
by $2\pi/|\mathbf{q}|$. The amplitude of the SDW and the energy gained by the system are determined by the product 
$\chi'_0(\mathbf{q})\;.\; V_q$. If $V_{\mathbf{q}}$ changes slowly in reciprocal space, a SDW will occur at the wavevector 
$\mathbf{q}$ for which $\chi'_0(\mathbf{q})$ is maximum, provided the mentioned criterion is satisfied. 
The imaginary part of the susceptibility, $\chi''_0{\mathbf{(q)}}$, gives information on the nesting properties. 
It is related to the so called nesting function 
$N_f(\mathbf{q})=\lim_{\omega->0}\frac{\chi''_0{\mathbf{(q)}}}{\omega}=\sum_{k}{\delta(\epsilon_F-\epsilon_{\mathbf{k}})\delta(\epsilon_F-\epsilon_{\mathbf{k+\mathbf{q}}})}$, where $\epsilon_F$ is the Fermi level.

\begin{figure}[!bth]
\centering
\includegraphics[width=8.3cm]{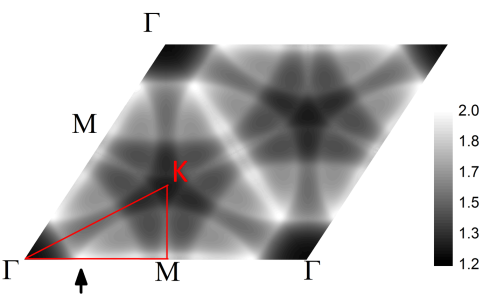}  
\caption{\label{fig:RealPartFullBZ} Real part of the static non interacting susceptibility $\chi'_0(\mathbf{q})$ of the NbS$_2$ monolayer (non SP) in the full first Brillouin zone, without SO. Values are in 1/eV. The arrow points out the absolute maximum, which is along the $\Gamma$M direction. The red line marks the edges of the irreducible Brillouin zone.}
\end{figure}

\begin{figure}[htb]
\centering
\includegraphics[width=8.3cm]{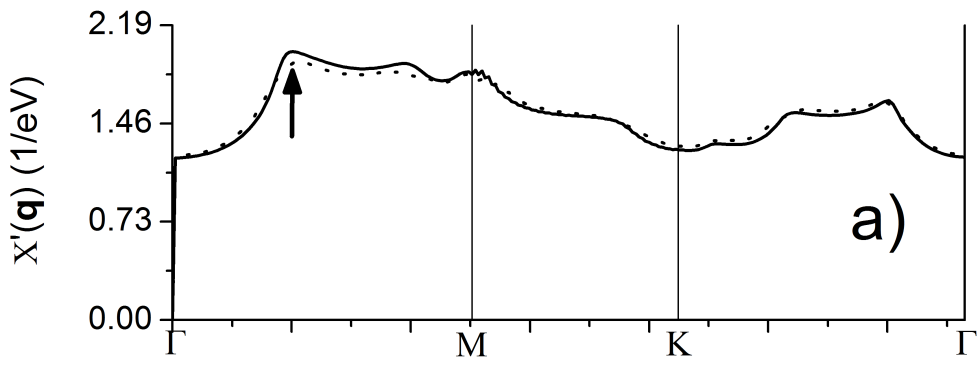} 
\includegraphics[width=8.3cm]{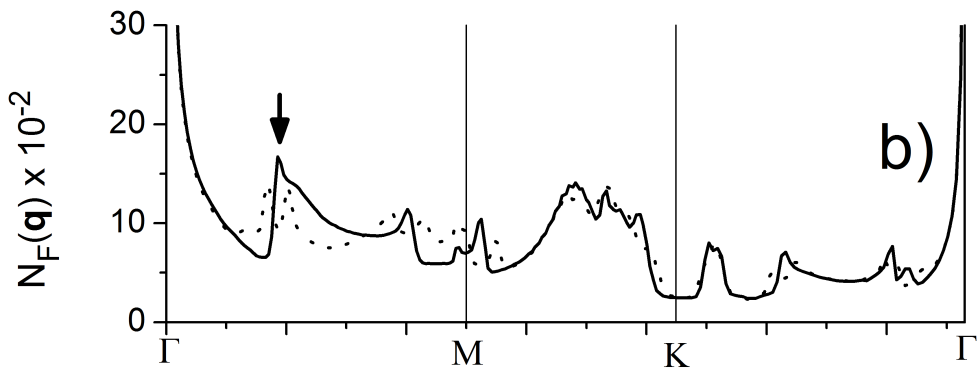} 
\caption{\label{fig:ChiProfiles} \textbf{a)} Real part of the static non interacting susceptibility, $\chi'_0(\mathbf{q})$ and \textbf{b)} nesting function N$_f(\mathbf{q})$ of the NbS$_2$ monolayer along the $\Gamma$MK$\Gamma$ path, with (dashed) and without (solid) SO. The arrows indicate the absolute maximum of $\chi'_0(\mathbf{q})$.}
\end{figure}

In Fig. \ref{fig:RealPartFullBZ} we show the calculated $\chi'_0(\mathbf{q})$ for the NbS$_2$ monolayer in the full first Brillouin zone.
The susceptibility has several local maxima for different $\mathbf{q}$ vectors. The highest maximun is found at $\mathbf{q_n}\approx0.2\mathbf{b_1}=(0.2,0)4\pi/\sqrt{3}a$ and its symmetry equivalent vectors. In Fig. \ref{fig:ChiProfiles}, we show 
the calculated $\chi'_0(\mathbf{q})$ and N$_f(\mathbf{q})$ along the high-symmetry directions  $\Gamma$-M-K-$\Gamma$. 
The most relevant feature to remark from this plot is that the nesting vector for which N$_f(\mathbf{q})$ is maximum
is $\mathbf{q_n}$. It matches with the one that maximizes $\chi'_0(\mathbf{q})$, both are indicated with arrows in the plot.
 We note that the calculated $\chi_0(\mathbf{q})$ depends weakly on the calculation settings and structural 
parameters being, in fact, very similar to the ones obtained for the undistorted NbSe$_2$ and TaSe$_2$ monolayers (without SO)\cite{YizhiGe2013}. When the SO coupling is included, the band crossing $\epsilon_F$ splits, giving rise to a concomitant splitting at $\mathbf{q_n}$ in $N_f$ and a slight broadening in $\chi'_0(\mathbf{q})$.

The direction of the magnetic wave pattern shown in Fig. \ref{fig:RigidRibbonMag}b is exactly the direction of $\mathbf{q_n}$, which is perpendicular to the ribbon edge (see Fig. \ref{fig:RigidRibbonMag}a). We estimate the wavelength of the resulting SDW from a least squares fit of the magnetic moments obtained for the NbS$_2$ nanoribbon of N=20 (by considering 12 innermost rows only). We get $\lambda=13.88$ \AA{}  and $\lambda=13.43$ \AA{}  for the unrelaxed and relaxed cases, respectively. Both values are very close (within 4\% and 8\% respectively) to the wavelength corresponding to $\mathbf{q_n}$. It is expected that for ribbons with larger widths, the outcoming $\lambda$'s should be even closer to $2\pi/|\mathbf{q_n|}$. The magnetic patterns of the monolayer (Fig. \ref{fig:2DPattern}a) should also approach this wavelength if a supercell larger than 4x4 is considered.

In view of this analysis we claim that the wave-like patterns obtained for the doped NbS$_2$ monolayers and ribbons are SDWs originated in the nesting of the 2D Fermi surface. In the presence of SDWs the NbS$_2$ monolayer remains metallic. The competing SDW phase presents a partial band splitting, while several dispersive bands still cross $\epsilon_F$.

\begin{figure}[thb]
\centering
\includegraphics[width=8.3cm]{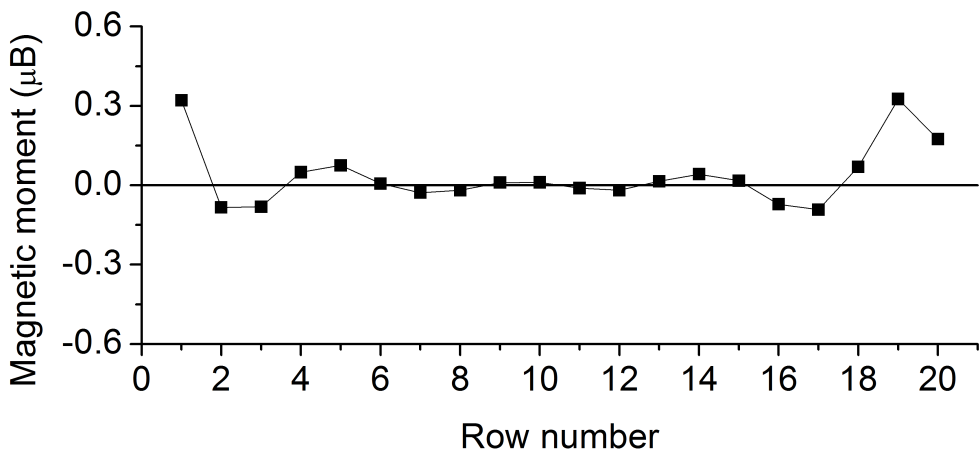} 
\caption{\label{fig:MagTaS2} Magnetic moments of the individual Ta atoms across the TaS$_2$ zig-zag ribbon of width N=20, unrelaxed (without SO). Rows of Ta atoms are numbered consecutively starting on the Ta terminated edge.}
\end{figure}

It is important to remark that besides the high suceptibility and ribbon edges, or other perturbations, a SDW will not be stable in a given system if $V_{\mathbf{q}}$ is not large enough. To show an example of such a situation, we study the
 magnetic configuration of the TaS$_2$ zig-zag ribbon with N=20. The exchange interaction in 4d metals (such as Nb) is stronger than in the 5d elements as Ta, due to the stronger spatial localization of the d orbitals with smaller quantum principal number. In Fig. \ref{fig:MagTaS2} we present the magnetic moments of the individual Ta atoms in the ribbon. Although there is a wave-like pattern, the oscillation decays rapidly towards the center. In this system the SDW is weaker. We alert the reader that the data of Fig. \ref{fig:MagTaS2} corresponds to an unrelaxed ribbon, without SO, merely used as a way to demonstrate the importance of the exchange interactions in the strength of SDW phases. The actual, physical TaS$_2$ zig-zag ribbon may be subject to CDW phases and stronger SO effects.

It is well known that DFT calculations in their local spin density approximation (LSDA) or in the gradient corrected GGA overestimate the tendency towards long range magnetic order in systems that are close to a quantum critical transition. 
The main reason behind this is the underestimation of quantum fluctuations in LSDA or GGA functionals, 
being these fluctuations particularly important in critical systems. 
Recently, an assessment of the magnitude of the spin-fluctuations beyond LSDA or GGA calculations has been made for a 
prototypical critical material as Pd \cite{larson2004}, and even more recently, for the ferromagnetic-paramagnetic 
transition of Ni$_3$Al under pressure \cite{ortenzi2012}. 
We propose that the NbS$_2$ monolayer is close to a quantum critical transition as well. 
A way to calculate the intensity of the zero-point spin fluctuations is through the fluctuation-dissipation theorem that reads,

\begin{equation}
{\xi}^2=\frac{2\hbar}{\Omega}\int d^3q \int \frac{d\omega}{2\pi}Im\chi(\mathbf{q},\omega)
\label{eq:fluc-dissip}
\end{equation}

where $\xi$ is the deviation of the spin density from its mean field value, $\Omega$ is the Brillouin zone area and $\chi(\mathbf{q},\omega)$ the dynamical susceptibility of the pristine NbS$_2$ monolayer.

Due to the complexity of getting the full dynamical susceptibility, in particular within a 
first-principles material specific approach, it is usual to do approximations.
One way to estimate $\xi$ is to consider in Eq. \ref{eq:fluc-dissip} the non-interacting 
$Im \chi_0(\mathbf{q},\omega)=\sum_{\bf{k}}[f(\epsilon_{\bf{k}})-f(\epsilon_{\bf{k+q}})]\delta(\epsilon_{\bf{k+q}}-\epsilon_{\bf{k}}-\hbar\omega)$. Here, $\epsilon_{\bf{k}}$ are the Kohn-Sham eigenvalues as before\cite{Note}.
When introducing $Im \chi_0(\mathbf{q},\omega)$ in Eq. \ref{eq:fluc-dissip}, we get $\xi$=0.2 $\mu_B$. 
This value is an underestimation of the quantum fluctuations of the system, since an enhancement of the spin-fluctuations is expected when considering the scattering of the electron-hole pairs 
due to all the electron-electron interactions \cite{Moriya}. In spite of this, the obtained value for $\xi$ is still
quite large, within the same order of magnitude as the calculated static magnetic moments of the magnetic configurations that compete in energy with the non magnetic one. We conclude that the spin-fluctuations can mitigate the long range magnetic instabilibilities in the pristine NbS$_2$ monolayer but, as shown before, they can be stabilized with doping, impurities, defects or edges.

It is worth noting that similar values for the static magnetic moments and the fluctuations ($\xi$) were obtained 
for Pd by using GGA for the exchange-correlation functional \cite{larson2004} and these fluctuations (paramagnons) 
were recently detected through inelastic neutron scattering experiments \cite{doubble2010}.

\section{Further discussion and conclusions\label{sec:Conclusions}}

In this work we have studied, by means of first principles calculations, the magnetic properties of the NbS$_2$ monolayer. 
We have shown that this system presents a high magnetic susceptibilty, large spin-fluctuations and 
it is on the verge of a SDW phase.  
We have also shown that the SDW states can be stabilized either by doping, defects, impurities or ribbon edges. 
The physical mechanism behind the SDWs is of pure electronic origin driven mainly by the nesting 
properties of this two-dimensional material and a rather strong Coulomb interaction in the 4$d$ band of the Nb atoms.

Even if no long range magnetic order has been experimentally observed in NbS$_2$, neither in bulk, nor thin films nor, 
in the monolayer, as far as we know, there are no reported experimental works investigating the spin-fluctuations of 
these systems. Since our results indicate that the monolayer is close to a quantum critical magnetic transition, 
these kind of measurements become highly desired.

On the other hand, bulk NbS$_2$ shows a superconducting phase below T=6 K with similar characteristics to its analogous 
2H-NbSe$_2$. In this last case, the system does show an ordered phase (CDW) whose connection  with the superconducting one is still under debate. Although there has been no suggestion  other than the electron-phonon coupling for the pairing mechanism of the superconductivity in the TMDC family\cite{Calandra2011,Leroux2012}, it is difficult to avoid making a conjecture regarding a possible connection of the SDW instabilities  predicted in this work and the superconductivity in NbS$_2$. However, this issue deserves further studies and it is beyond the scope of this paper.

\begin{acknowledgments}

The authors belong to the Institute of Nanoscience
and Nanotechnology (INN) of the Atomic Energy Agency (CNEA), Argentina.
They acknowledge financial support from ANPCyT (PICT-2011-1187) and CONICET (PIP00069).

F.G. and V.L.V. contributed equally to this work.

\end{acknowledgments}


\begin{thebibliography}{99}
\bibitem{WIEN2kspecific} WIEN2k specific settings: the muffin tin radii used were 1.2 au for S and 2.0 au for Nb and Ta. Plane wave cutoff: RKMAX = 8. Energy and charge distance cutoff for self consistent iterations: $10^{-4}$ Ry and $10^{-4}$ e, respectively.
\bibitem{NotaZC} The difference in the value of the $\mathbf{zc}$ parameter is most likely due to the van der Waals interactions between layers, which are absent when a single monolayer is considered, but are present in the bulk.
\bibitem{Note} The calculated $\chi_0$ is not exactly \emph{non-interacting} because we use the GGA eigenvalues. It is usually called like this in the literature and we continue with this terminology.
\end{thebibliography}
\end{document}